\documentclass[11pt]{article}
\usepackage{a4wide}
\usepackage[pdftex]{graphicx}
\usepackage{algorithm}
\usepackage[noend]{algpseudocode}

\usepackage{url}
\usepackage[colorlinks,linkcolor=black,citecolor=black,filecolor=black,urlcolor=black]{hyperref}
\usepackage[all]{hypcap}
\usepackage{xcolor}
\usepackage{multirow,multicol}
\usepackage{caption}
\usepackage{subfigure}
\usepackage{listings}
\usepackage[fleqn]{amsmath}
\usepackage{amssymb}
\usepackage{stmaryrd}
\usepackage{comment}
\usepackage{float}
\usepackage{rotating}
\usepackage{booktabs}

\usepackage{xpatch}
\usepackage[frozencache=true,cachedir=.]{minted}

\makeatletter
\AtBeginEnvironment{minted}{\dontdofcolorbox}
\def\dontdofcolorbox{\renewcommand\fcolorbox[4][]{##4}}
\xpatchcmd{\inputminted}{\minted@fvset}{\minted@fvset\dontdofcolorbox}{}{}
\xpatchcmd{\mintinline}{\minted@fvset}{\minted@fvset\dontdofcolorbox}{}{} 
\makeatother
\usepackage{csvsimple}
\usepackage{wrapfig}

\graphicspath{ {images/} }

\usepackage{pifont}
\newcommand{\cmark}{\textrm{\ding{51}}}
\newcommand{\xmark}{\textrm{\ding{55}}}

\algblockdefx[CONSTANT]{Constant}{EndConstant}%
  [1]{\textbf{constant} #1}
  
\algblockdefx[MATCH]{Match}{EndMatch}%
  [1]{\textbf{match} #1}
  
\algblockdefx[CASE]{Case}{EndCase}%
  [1]{\textbf{case} #1}

\newmintinline{java}{}

\graphicspath{ {images/} }

\usepackage{authblk}

\title{IFDS Taint Analysis with Access Paths}
\author{Nicholas Allen, Fran\c{c}ois Gauthier, Alexander Jordan}
\affil{Oracle Labs}
\affil{first.last@oracle.com}
\begin{document}
\maketitle

\section*{Abstract}

Over the years, static taint analysis emerged as 
the analysis of choice to detect some of the most common web application
vulnerabilities, such as SQL injection (SQLi) and cross-site scripting (XSS)~\cite{OWASP}. Furthermore,
from an implementation perspective, the IFDS dataflow framework~\cite{reps1995precise}
stood out as one of the most successful vehicles to implement static taint analysis 
for real-world Java applications~\cite{tripp2009taj, tripp2013andromeda, 
arzt2014flowdroid}.

While existing approaches scale reasonably to medium-size applications (e.g. 
up to one hour analysis time for less than 100K lines of code), our experience 
suggests that no existing solution can scale to very large industrial code bases 
(e.g. more than 1M lines of code). In this paper, we present our novel IFDS-based 
solution to perform fast and precise static taint analysis of very large industrial 
Java web applications.

Similar to state-of-the-art approaches to taint analysis, our IFDS-based taint 
analysis uses \textit{access paths} to abstract objects and fields in a program. 
However, contrary to existing approaches, our analysis is demand-driven, 
which restricts the amount of code to be analyzed, and does not rely on a 
computationally expensive alias analysis, thereby significantly improving scalability.

\section{Background}


The IFDS analysis framework is a dataflow analysis framework for solving
inter-procedural, finite, distributive, subset (IFDS) problems. Flow functions $f$
are defined over a finite domain of dataflow facts $D$, and have to be distributive
over the meet operator, union (i.e. $f(a) \cup f(b) = f(a \cup b)$). These flow
functions are defined by the specific analysis (in our case, taint analysis), to
specify the effect on dataflow facts that corresponds with the execution of the 
statement at the given program point. The IFDS analysis framework solves dataflow
problems efficiently by reducing them to graph reachability problems.
The reachability of a particular node in the graph represents whether a
particular dataflow fact holds at a particular program point.

There are two main variants of the IFDS analysis algorithm. The forward version
of the analysis propagates facts forward through the program and exhaustively
computes the dataflow facts that hold at each program point. In contrast, the backward
version of the analysis is demand-driven, i.e., whether a
particular fact holds at a particular program point is determined \emph{on-demand}, in
response to client queries. Our work is based on the backward version of the
IFDS analysis algorithm.

The IFDS analysis framework achieves efficient inter-procedural analysis via
function summarisation. Function summaries are generated on demand during the
analysis, and represent the backward reachability from an end fact to a
set of start facts. There can be multiple summaries generated for a single
function, one for each relevant end fact.

\begin{figure}
\centering
\includegraphics[width=\textwidth]{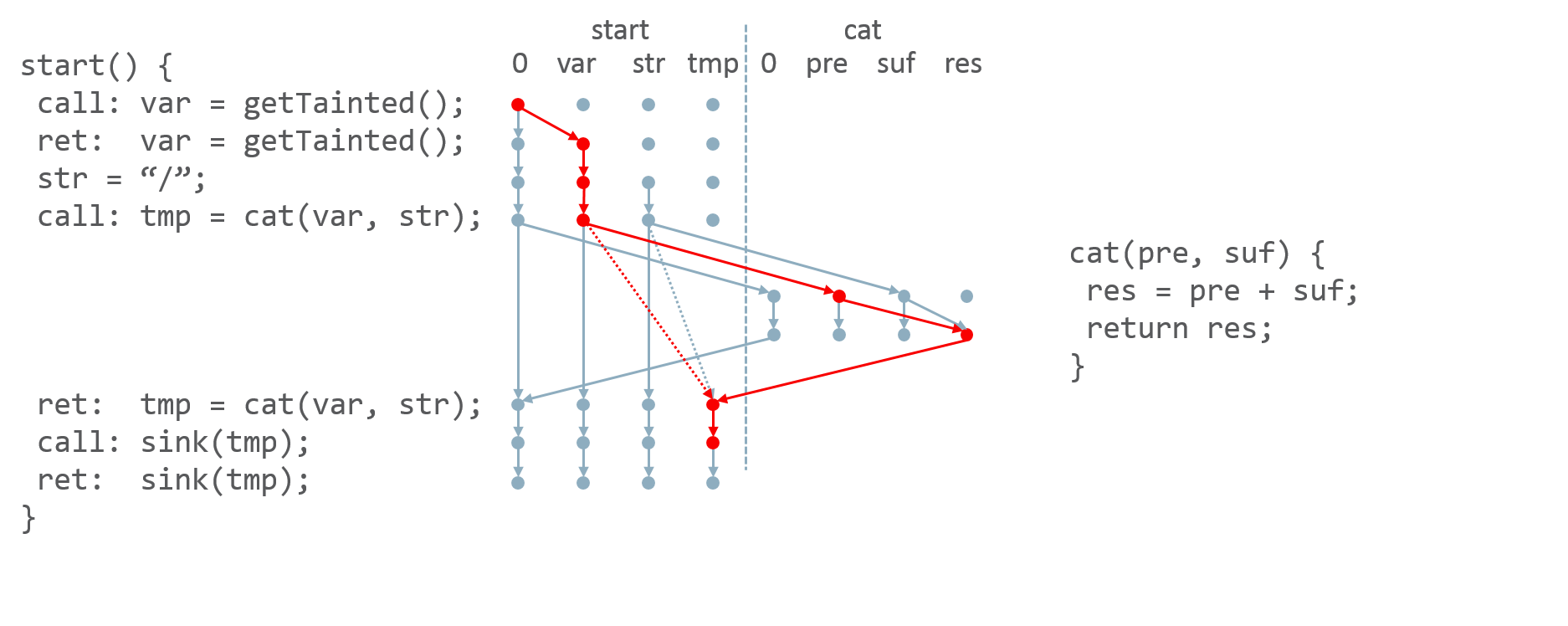}
\caption{Simple IFDS Taint Analysis Example}
\label{fig:IFDSTaintExample}
\end{figure}

IFDS operates on the \textit{exploded supergraph} of a program, which is
an inter-procedural control-flow graph (ICFG) where each node in the ICFG is
exploded into as many nodes as there are dataflow facts. Figure \ref{fig:IFDSTaintExample} 
shows an example of an IFDS analysis (in this case, a simple taint analysis)
applied to the following program:

\begin{minted}[fontsize=\footnotesize,
               linenos,
               gobble=2,
               frame=lines,
               framesep=2mm]{java}
  void start() {
    String var = getTainted();
    String str = "/";
    String tmp = cat(var, str);
    sink(tmp);
  }
  String cat(String pre, String suf) {
    String res = pre + suf;
    return res;
  }
\end{minted}

A particular dataflow fact holds at a particular statement if and only if the
associated node in the exploded supergraph is reachable. Edges in the exploded
supergraph encode the flow functions. In Figure~\ref{fig:IFDSTaintExample}, the
fact that the sink node (\javainline{sink(tmp)}, $tmp$) is reachable from the
entry node (\javainline{var = getTainted()}, \textbf{0}) indicates the existence
of an execution path, highlighted in red, where tainted data reaches a
security-sensitive sink.

In our approach, the exploded supergraph is extracted from programs in SSA form,
and we use IFDS to encode a taint analysis over k-bounded access paths.

\begin{description}
\item [SSA Form] We use \textit{Static Single Assignment} (SSA)
  form~\cite{Cytron1989} as an intermediate representation for our analysis.
  SSA requires that each use of a variable has a single definition. If there
  are multiple definitions for a use in the original program, a
  $\phi$-function (or $\phi$-node) is inserted in the control flow graph (CFG).
  \item [Access Paths] Our analysis propagates \textit{access paths} of the form 
    $x.f.g$, where $x$ is a local variable, and $f$ and $g$ are fields.
    Specifically, $x.f.g$ represents the value that is retrieved by first dereferencing $x$ in the 
    current scope and then dereferencing fields $f$ and $g$ from the heap. Because
    access paths are unbounded, we use \textit{k-limiting} to bound their length
    to a pre-defined length $k$. When an access path reaches a length of $k$, further
    appends are simply ignored (i.e. access paths longer than \textit{k} are
    assumed to be untainted). Our approach uses a default value of $k=5$.
  \item [Taint Analysis] The goal of taint analysis is to find and report
    dataflow from \emph{sources} to security-sensitive \emph{sinks} that does not
    undergo a \emph{sanitisation} operation. In general, \emph{taint labels}
    are used to track complementary taint information (e.g. SQLi vs. XSS).  For
    the sake of simplicity, in this paper, we assume that values are either 
    \emph{tainted} or \emph{untainted}, and that there are no sanitisation 
    functions. Extending our approach to support sanitisation and taint labels
    is straightforward, and our implementation does support both of these
    features.
\end{description}

\section{Approach} \label{approach}

Our analysis performs a demand-driven, backward taint analysis. 
Similar to recent work on IFDS-based static taint
analysis~\cite{tripp2013andromeda, arzt2014flowdroid}, our analysis tracks
taint through objects and fields by propagating \textit{access paths}.

In our implementation, we adapted the extended forward IFDS algorithm presented
by Naeem et al. in~\cite{naeem2010practical} to the on-demand backward analysis
presented by Reps et al. in~\cite{reps1994}. Our extended algorithm adds the
following three optimisations described in~\cite{naeem2010practical} to the
original backward IFDS algorithm:

\begin{enumerate}
  \item Lazy computation on the exploded 
    supergraph (e.g. the graph with one node per instruction and dataflow fact, 
    as shown in Figure~\ref{fig:IFDSTaintExample}).
    Lazy computation of the exploded supergraph reduces the memory footprint
    of the analysis by ensuring that only the relevant portions of the exploded
    supergraph are built.
  \item Support for $\phi$-nodes in SSA form.
    Because $\phi$-nodes at merge points in the CFG cause a loss of precision
    during dataflow analysis, this extension ensures that the IFDS algorithm delays merging
    of dataflows until \textit{after} $\phi$-nodes have been processed.
  \item Providing the procedure-call flow
    function (in~\cite{naeem2010practical} this was applied to procedure-return
    in the forward version of the analysis) with information about the
    caller-side state from the time of the procedure-return, allowing the
    callee-side state to be mapped to the caller-side context more precisely.
\end{enumerate}

We now present the intra-procedural flow functions that define our backward IFDS
taint analysis. Algorithm~\ref{alg:FlowAlgorithm} defines (in the \textsc{Flow}
procedure) the flow functions for allocation, assignment, field-load and
field-store statements in a Java program.
For each type of statement ($\llbracket {stmt.} \rrbracket$) the flow function
defines which facts must hold (any) \emph{before}
the statement, for a given fact to hold \emph{after} the execution of the
statement.
And because our analysis operates on access paths, a flow function maps an access path of
the form $b.f_1 \ldots f_n$, where $b$ is the base variable, and $f_1 \ldots
f_n$ is a sequence of fields, to a set of access paths.
Note that we omit the inter-procedural call and return flow functions, because
they simply convert arguments and return values between callers and callees
without modifying access paths. The \textsc{Flow} procedure is invoked during
the execution of our backward IFDS algorithm implementation as statements are
processed, to perform on-the-fly exploded supergraph construction for
intra-procedural dataflow edges.

\newcounter{algcasecounter}
\newcommand*{\algcaselabel}[1]{\refstepcounter{algcasecounter} \label{#1}
\Comment {$(\ref{#1})$}}

\begin{algorithm}
\begin{algorithmic}
\caption{Intra-procedural flow functions}
\label{alg:FlowAlgorithm}
\Constant{k}
\EndConstant
\Procedure{Flow}{$statement, (b.f_1 \ldots f_n)$}
  \Match {$statement$}
    \Case {$\llbracket x = \textbf{new} \rrbracket$}
    \algcaselabel{algcase:alloc}
      \If {$x = b$} 
        \Return {$\emptyset$}
      \Else
        $ $ \Return {$\{( b.f_1 \ldots f_n )\}$}
      \EndIf
    \EndCase
    \Case {$\llbracket x = y \rrbracket$} \algcaselabel{algcase:assign}
      \If {$x = b$}
        \Return {$\{( y.f_1 \ldots f_n )\}$}
      \Else
        $ $ \Return {$\{( b.f_1 \ldots f_n )\}$}
      \EndIf
    \EndCase
    \Case {$\llbracket x = Taint Source() \rrbracket$}
    \algcaselabel{algcase:taint}
      \If {$x = b$}
        \Return {$\{\mathbf{0}\}$}
      \Else
        $ $ \Return {$\{( b.f_1 \ldots f_n )\}$}
      \EndIf
    \EndCase
    \Case {$\llbracket x = y.g \rrbracket$} \algcaselabel{algcase:load}
      \If {$x = b$}
        \State {$z.g_1 \ldots g_m \gets $} \Call{Reify}{$(y.g)$}
        \If {$m + n > k$}
          \Return {$\emptyset$}
        \Else
          $ $ \Return {$\{( z.g_1 \ldots g_m.f_1 \ldots f_n )\}$}
        \EndIf
      \Else
        $ $ \Return {$\{( b.f_1 \ldots f_n )\}$}
      \EndIf
    \EndCase
    \Case {$\llbracket x.g = y \rrbracket$} \algcaselabel{algcase:store}
      \State {$z.g_1 \ldots g_m \gets $} \Call{Reify}{$(x.g)$}
      \If {$z = b \textbf{ and } m \leq n \textbf{ and } g_1 \ldots g_m = f_1
      \ldots f_m$}
        \If {$\forall i \in [1,m] \text{, } g_i \text{ is not an array}$}
          \Return {$\{( y.f_{m+1} \ldots f_n )\}$}
        \Else
          $ $ \Return {$\{( y.f_{m+1} \ldots f_n ) \text{,} ( b.f_1 \ldots f_n
          )\}$}
        \EndIf
      \Else
        $ $ \Return {$\{( b.f_1 \ldots f_n )\}$}
      \EndIf
    \EndCase
  \EndMatch
\EndProcedure

\Procedure{Reify}{$(b.f_1 \ldots f_n)$}
  \Match \Call{Definition}{$b$}
    \Case {$\llbracket b = y.g \rrbracket$}
      \State \Return \Call{Reify}{$(y.g.f_1 \ldots f_n)$}
    \EndCase
    \Case {$\mathbf{default}$}
      \State \Return {$(b.f_1 \ldots f_n)$}
    \EndCase
  \EndMatch
\EndProcedure
\end{algorithmic}
\end{algorithm}

%
%
%
%

Case~\ref{algcase:alloc} defines the flow function for allocation statements.
The incoming access path is mapped to the empty set ($\emptyset$) if its base variable $b$ matches the newly assigned local variable $x$ to capture the
fact that access paths rooted at $x$ cannot exist before $x$ is allocated. Otherwise, the identity 
function is applied. Case~\ref{algcase:assign} defines the flow function for
assignments of the form $x = y$. The base variable $b$ of the incoming access path is replaced with $y$ if $b$ matches $x$. 
Case~\ref{algcase:taint} defines the flow function for assignment of tainted
values.
If $b$ matches $x$, the incoming access path is mapped to the null fact
($\mathbf{0}$), to capture the fact that $x$ became tainted at that specific
point in the program. 

%
%
%

Cases~\ref{algcase:load} and \ref{algcase:store} define the flow functions for
loads of the form $x = y.g$ and stores of the form $x.g = y$, respectively.
Because our algorithm works with programs in the SSA intermediate representation
(IR), care must be taken to reify statements involving multiple stores and
loads. Indeed, translation to an IR usually deconstructs field accesses into
multiple sub-statements using temporary variables that require reification
before analysis. To address this issue, our analysis performs an on-demand,
intra-procedural reification step (the \textsc{Reify} procedure) before
processing any store or load instruction, which determines the full access path
referenced by the load or store statement. 

Hence, Case~\ref{algcase:load} defines the flow function for loads of the
post-reification form $x = z.g_1 \ldots g_m$. The base variable $b$ is replaced
with $z$, and the loaded fields $g_1 \ldots g_m$ are prepended to the access
path if $b$ matches $x$ (unless the length of the new access path exceeds the
pre-defined limit \textit{k}, in which case the empty set is returned).
Case~\ref{algcase:store} defines the flow function for stores of the
post-reification form $z.g_1 \ldots g_m = y$. The base variable $b$ is replaced
with $y$, and fields $f_1 \ldots f_m$ are removed from the incoming access path
if $b$ matches $z$ and the stored fields $g_1 \dots g_m$ match $f_1 \ldots f_m$
(i.e. the stored fields form a prefix of the incoming access path). If any of
the stored fields is an array, the incoming access is also preserved because our
analysis is array-insensitive (e.g. it cannot reason about the exact array cell
that is loaded), and hence cannot invalidate the incoming access path.

\begin{wrapfigure}{r}{0.2\textwidth}
\begin{center}
\begin{minted}[fontsize=\footnotesize,
               gobble=2,
               frame=lines,
               framesep=2mm]{java}
  tmp1 = y.f;
  tmp2 = tmp1.g
  tmp2.h = a;
\end{minted}
\end{center}
\end{wrapfigure}

We now explain the reification step (the \textsc{Reify} procedure) in more
detail by way of an example. Assume that $a$ is tainted, and that we are
computing the flow function of the incoming access path $y.f.g.h$ and the
statement \javainline{tmp2.h = a}. Without reification, Case~\ref{algcase:store}
would wrongly conclude that \javainline{tmp2.h = a} has no impact on $y.f.g.h$
(as the base variables, $tmp2$ and $y$, do not match). To determine that store
to $tmp2.h$ does, in fact, affect $y.f.g.h$, the reification step starts by
tracking the definition of the base variable of the store/load. Then, if the
definition is a load statement, the reification step replaces the base variable
of the original store/load with the loaded access path, and starts tracking the
definition of the base variable of the loaded access path. This is done
recursively until it reaches a definition that is not a load statement. Once the
reification step completes, the appropriate flow function can be applied to the
reified store/load statement.
 
In our example, when processing the statement \javainline{tmp2.h = a}, the reification step 
would start by tracking the definition of the base variable $tmp2$. Then, it would replace 
$tmp2.h$ with $tmp1.g.h$, and start tracking the definition of $tmp1$. Finally, it would
replace $tmp1.g.h$ with $y.f.g.h$. Thus, when the flow function defined in
Case~\ref{algcase:store} is applied to the reified statement \javainline{y.f.g.h
= a}, it correctly propagates taintedness from $a$ to $y.f.g.h$.


\subsection{Working Example}

\begin{figure}
\begin{minted}[fontsize=\footnotesize,
               linenos,
               gobble=2,
               frame=lines,
               framesep=2mm]{java}

  public class Box {
    private String f;

    public void put(String str) {     // 14. summary(this.f) = {arg0}
      this.f = str;                   // 13. str
    }                                 // 12. this.f

    public String get() {             // 6. summary(<ret>) = {this.f}
      String str = this.f;            // 5. this.f
      return str;                     // 4. str
    }                                 // 3. <ret>
  }

  public static Box copy(Box box) {   // 19. summary(<ret>.f) = {arg0.f}
    Box cpy = new Box();              // 18. box.f
                                      // 17. box.f
    String data = box.get();          // 16. reuse summary(Box.get, <ret>)
                                      // 15. data
    cpy.put(data);                    // 11. compute summary(Box.put, this.f)
    return cpy;                       // 10. cpy.f
  }                                   // 9. <ret>.f

  public static void foo() {
    String tainted = getTainted();    // 24. 0 (null fact)
    Box box1 = new Box();             // 23. tainted
                                      // 22. tainted
    box1.put(tainted);                // 21. reuse summary(Box.put, this.f)
                                      // 20. box1.f
    Box box2 = copy(box1);            // 8. compute summary(copy, <ret>.f)
                                      // 7. box2.f
    String boxData = box2.get();      // 2. compute summary(Box.get, <ret>)
    sink(boxData);                    // 1. boxData
  }

\end{minted}
\caption{A simple program annotated with dataflow facts, as propagated by our algorithm. Numbers in the comment show the order in which statements are processed.}
\label{example-ifds}
\end{figure}

Figure~\ref{example-ifds} demonstrates our approach applied to an example
program. In this example, the $foo$ method obtains tainted data from the
$getTainted$ method (the taint source), stores it into the field of a $Box$
object ($box1$), then makes a copy of that object ($box2$), retrieves the data
stored in the field of the copy and passes it to the $sink$ method (the taint
sink).

To determine whether the data passed in at the sink, i.e. the $boxData$
variable, is tainted, the analysis works backward from the sink (line 32),
tracking the dataflow fact $boxData$.
In the prior statement (line 31), $boxData$ is assigned the return value of
Box.get, so the summary for Box.get for the return value must be computed.
The summarisation of Box.get starts at the return statement (line 10), tracking
the dataflow fact $str$ (the returned variable). In the prior statement (line
9), Case~\ref{algcase:load} applies, which maps back to the fact $this.f$,
after which the method entry has been reached, so the summary generated for
Box.get establishes flow to the return value from $this.f$. Returning to the call to
Box.get (line 31), the summary is applied (substituting the actual $this$ object
passed in), resulting in a transfer to the dataflow fact $box.f$.
In the prior statement (line 29), $box2$ is assigned the return value of $copy$,
so the summary for that method for the field $f$ of the return value must be
generated.

This summarisation of the $copy$ method starts at the return (line 20) with the
dataflow fact $cpy.f$. In the prior statement (line 19), the $Box.put$ method is
invoked on $cpy$, so the effect of method $Box.put$ on $this.f$ must be summarised.
Case~\ref{algcase:store} is applied for the store statement in $Box.put$ and
this summary produced is that $this.f$ flows from the argument $str$.

Returning to $copy$ (line 19), the summary is applied, transferring to the
fact $data$ that is assigned the return of $Box.get$ in the prior statement
(line 17). The already-computed summary for $Box.get$ is applied, transferring
to $box.f$, which is unaffected by the prior new statement, so the summary
generated for $copy$ is that the field $f$ of the return value flows from the
field $f$ of the argument.

Returning to $foo$ (line 29), the summary for $copy$ is applied, transferring
from $box2.f$ to $box1.f$. At the prior statement (line 27), $Box.put$ is
invoked on $box1$. The already computed summary for $Box.put$ is applied,
transferring to $tainted$, which is unchanged until it is assigned the return
value of $getTainted$ (line 24). For this example, $getTainted$ is designated as
a taint source, so the dataflow fact $tainted$ maps to the null fact.

The null fact always holds, and the analysis has demonstrated that the $boxData$
fact at the sink is reachable from the null fact. Therefore, the $boxData$
variable passed to the sink is tainted, and so a bug would be reported for this
example.

\section{Initial Results}

\begin{table}[bht]
  \centering
  \begin{tabular}{l@{\hspace{1em}}rrrrc@{\hspace{2em}}rrrrc}
    \toprule
    \multirow{2}{*}{Benchmark} & \multicolumn{5}{c}{Legacy Taint Analysis} & \multicolumn{5}{c}{IFDS-AP Taint Analysis} \\
    & TP & TN & FP & FN & Runtime & TP & TN & FP & FN & Runtime\\
    \midrule
    \begin{filecontents*}{data/benchmarks.csv}
Benchmark,TP,TN,FP,FN,Runtime,ifdsTP,ifdsTN,ifdsFP,ifdsFN,ifdsRuntime
Securibench,99,0,10,39,0.1568,101,0,12,37,0.0184
WebGoat,35,0,1,33,0.5912,35,0,3,33,0.504
OWASP,501,533,451,324,3.7784,732,353,772,100,2.732
\end{filecontents*}
\csvreader[head to column names, late after line=\\]{data/benchmarks.csv}{}%
    {\Benchmark & \TP & \TN & \FP & \FN & \Runtime & \ifdsTP & \ifdsTN & \ifdsFP & \ifdsFN & \ifdsRuntime}
    \bottomrule
  \end{tabular}
  \caption{Results for analysis benchmarks}
  \label{tab:bench}
\end{table}
\vspace{2em}
\begin{table}[bht]
  \centering
  \begin{tabular}{lrrrc}
    \toprule
    Taint Analysis & TP & FP & Unknown & Runtime\\
    \midrule
    \begin{filecontents*}{data/product.csv}
Analysis,TP,FP,Unknown,Runtime
Legacy,96,9,9,6m15s
IFDS-AP,121,14,41,3m3s
\end{filecontents*}
\csvreader[head to column names, late after line=\\]{data/product.csv}{}%
    {\Analysis & \TP & \FP & \Unknown & \Runtime}
    \bottomrule
  \end{tabular}
  \caption{Results for Oracle product A}
  \label{tab:prod}
\end{table}

Our technique has been implemented in the context of Parfait~\cite{parfait},
and applied to the task of detecting security vulnerabilities in Java EE web
applications, such as SQL injections, cross-site-scripting, etc.

Table~\ref{tab:bench} shows the results of our analysis applied to three
analysis benchmarks, Securibench, WebGoat and OWASP, compared with the results
of our previously used taint analysis. Table~\ref{tab:prod} shows the
comparison with the analysis applied to an Oracle product. The results
demonstrate that our IFDS analysis detects more real bugs, and does so with
significantly reduced runtime. 

\section{Related Work}

Static taint analysis for the detection of vulnerabilities has a long history
in the research community. In this section, we describe and contrast the 
state-of-the-art approaches that are most closely related to ours.

In~\cite{livshits2005finding}, authors propose to use a
flow-insensitive points-to analysis to support a client taint analysis for
Java Enterprise Edition (JEE) web applications. While more modern approaches
gained in scalability and precision, this paper was seminal and triggered a
lot of follow-up research, as presented below.

TAJ~\cite{tripp2009taj} used an approach called \textit{thin
slicing} to perform taint analysis of JEE web applications. In thin slicing,
IFDS is used for flow-sensitive reasoning about tainted flows through local
variables while flows through the heap are handled by using flow-insensitive
pre-computed points-to information. TAJ propagates taint information in a
forward manner, from sources to sinks. Furthermore, TAJ bounds its analysis
using various heuristics to keep its runtime and memory usage to acceptable
levels.

Andromeda~\cite{tripp2013andromeda} first introduced the idea
of using access paths in an IFDS-based setup to compute alias and taint analysis
of JEE web applications simultaneously. In Andromeda, tainted access paths are
propagated in a forward manner, from entry points of the program to
security-sensitive sinks. Moreover, Andromeda also computes on-demand aliasing
by launching a backward alias analysis whenever the forward analysis reaches an
assignment to a field. Then, the forward taint analysis propagates taint through
the newly discovered aliases, and so on until a fixed point is reached. Because
the length of access paths is unbounded, Andromeda limits their length to a
user-specified value \textit{k}, a process known as \textit{k-limiting}.

FlowDroid~\cite{arzt2014flowdroid} integrated the dataflow
equations of Andromeda into the IFDS framework and improved precision by
sharing information between the taint and alias analyses. Indeed, in FlowDroid,
aliases become tainted only \textit{after} the original access path becomes
tainted, a mechanism referred to as ``activation statements'' in the paper. Furthermore,
FlowDroid includes support for Android-specific framework constructs that are
hard to analyse statically. FlowDroid also uses a forward taint analysis combined
with an on-demand backward alias analysis that uses \textit{k-limiting}.

Boomerang~\cite{spath2016boomerang} generalised the IFDS-based
alias analysis of FlowDroid and decoupled it from the taint analysis. In Boomerang,
client analyses can issue alias queries that will be solved on-demand. Boomerang
also extends previous work by replacing access paths with access graphs that can
represent multiple access paths of indefinite length. Otherwise, Boomerang reuses
the forward and backward dataflow equations introduced in Andromeda~\cite{tripp2013andromeda}
to compute alias information. When using Boomerang instead of 
its original alias analysis, FlowDroid could analyze more applications in a
given timeout.

\section{Comparison with Existing Approaches}


An important component of our analysis is the ability to report bug traces for
the analysis results. Indeed, several static program analyses do not keep track of
\textit{provenance} information and hence cannot produce explanations in the form
of a bug trace from a sink to a source. Because IFDS is fully flow- and
context-sensitive and because it stores provenance information in the form of \textit{Path Edges},
our approach naturally produces understandable bug traces out-of-the-box.  


Furthermore, contrary to existing approaches that are geared towards soundness, our taint analysis 
implementation is geared towards high scalability. For example, in
our implementation, we use the \textit{flyweight}~\cite{Gamma1995} design pattern to
ensure that each access path is created only once in memory and reused as many times
as needed. Furthermore, we also
optimise away nodes in the exploded supergraph that have only one predecessor and
for which the transfer function is the identity function. Because most nodes fall
in this category (e.g. most statements have only one predecessor and don't
modify tainted access paths), this optimisation speeds up the analysis significantly. (We
observed a speedup of up to 45\% on large programs).

Moreover, contrary to~\cite{tripp2013andromeda, arzt2014flowdroid} our taint
analysis deliberately omits computing complete aliasing information, which would require an interplay 
between our backward taint analysis and a forward alias propagation analysis. This 
deliberate trade-off of soundness for scalability drastically reduces the theoretical
complexity of our algorithm. Precisely, according to the definitions in~\cite{reps1995precise},
our analysis is h-sparse, because, as shown in section~\ref{approach},
every transfer flow function produces at most 2 facts, and $2 << |D|$, where $|D|$ is
the cardinality of the dataflow domain (e.g. all possible access paths in a program). 
According to~\cite{reps1995precise}, \textit{h-sparse} problems have a complexity of $O(Call~D^3 + hED^2)$,
where $Call$ is the number of call sites, $D$ is the dataflow domain, and $E$ is
the set of intra-procedural edges. On the other hand, because the number of aliases of a given 
variable cannot be bounded to $h << |D|$ for non-trivial programs, an analysis that 
computes taint analysis together with an alias analysis is said to be \textit{Distributive}
and has a complexity of $O(ED^3)$. 

While our technique and \cite{tripp2013andromeda} both limit access paths to a
maximum size of \textit{k}, the approach used in \cite{tripp2013andromeda}
favours soundness by appending a Kleene star to access paths exceeding
\textit{k}, that are considered to match all other access paths sharing the
same \textit{k}-prefix. This may result in spurious taint flows being explored.
In contrast, our \textit{k-limiting} approach favours precision (and hence
scalability, as fewer potential taint flows are explored), by ignoring any taint
flows involving access paths exceeding \textit{k}.

Table~\ref{tab:novelty} summarizes the novelty of our approach with respect to 
state-of-the-art approaches.

\begin{table}[bht]
  \centering
  \begin{tabular}{|l|l|l|l|l|}
    \toprule
    Approach & Full alias analysis & Complete bug trace & Scalability &
    Complexity
    \\
    \toprule
    TAJ & $\cmark$ & $\xmark$ & High (Bounded) & $O(ED^{3})$\\
    \midrule
    Andromeda & $\cmark$ & $\cmark$ & Medium & $O(ED^{3})$ \\
    \midrule
    FlowDroid & $\cmark$ & $\cmark$ & Low & $O(ED^{3})$\\
    \midrule
    Our technique & $\xmark$ & $\cmark$ & Very high & $O(Call D^{3} +
    2ED^{2})$\\
    \bottomrule
  \end{tabular}
  \caption{Novelty of our approach}
  \label{tab:novelty}  
\end{table}

\bibliographystyle{alpha}
\bibliography{biblio}
\end{document}